\input{aipcheck}

\documentclass[
    ,final            % use final for the camera ready runs
  ]
  {aipproc}

\layoutstyle{6x9}
\usepackage{amssymb}
\begin{document}

\title{$\alpha_s$ from the Lattice and Hadronic $\tau$ Decays}



\author{Kim Maltman}{
  address={Math and Stats, York Univ., 4700 Keele St., Toronto, ON CANADA
M3J 1P3}
  ,altaddress={CSSM, Univ. of Adelaide, Adelaide, 5005 SA, Australia} % additional visiting address
}

\begin{abstract}
Until recently, determinations of $\alpha_s(M_Z)$ from
hadronic $\tau$ decays and the analysis of short-distance-sensitive
lattice observables yielded results which, though precise, were not in good
agreement. I review new analyses that bring these into good
agreement and provide some details on the source of the main
changes in the $\tau$ decay analysis.
\end{abstract}

\maketitle

%%%%%%%%%%%%%%%%%%%%%%%%%%%%%%%%%%%%%%%%%%%%
%% MAINMATTER
%%%%%%%%%%%%%%%%%%%%%%%%%%%%%%%%%%%%%%%%%%%%

%%%%%%%%%%%%%%%%%%%%%%%%%%%%%%%%%%%%%%%%%%%%
%% Sample figure:
%%
%% The option [height=...] scales the picture to the given height,
%% without it it would be printed at its nominal size
%%%%%%%%%%%%%%%%%%%%%%%%%%%%%%%%%%%%%%%%%%%%

Until recently the determinations of
$\alpha_s(M_Z)$ from (i) perturbative analyses of short-distance-sensitive
lattice observables (yielding $0.1170(12)$~\cite{latticealphas05}), and
(ii) from finite energy sum rule (FESR) analyses involving
hadronic $\tau$ decay data (yielding $0.1212(11)$~\cite{davieretal08}),
both claiming high precision, produced central values differing from
one another by $\sim 3\sigma$. In the past year, significant updates to both
analyses have appeared, bringing the two determinations into
excellent agreement. We outline the important features of these updates 
responsible for this change in what follows.

The original lattice determination~\cite{latticealphas05}, employed
a number of lattice observables, $O_k$, and the perturbative $D=0$
expansions for these observables, 
\begin{equation}
O_k=\sum_{N=1}{\overline{c}}_N^{(k)}[\alpha_V(Q_k)]^N
\equiv D_k\alpha_V(Q_k)\sum_{M=0}c_M^{(k)}[\alpha_V(Q_k)]^M
\label{3loopPT}\end{equation}
where $\alpha_V$ is a coupling defined in
Refs.~\cite{latticealphas05,hpqcd08}, $Q_k=d_k/a$ is the BLM scale
for $O_k$, and the ${\overline{c}}_{1,2,3}^{(k)}$ 
(equivalently, $D_k, c_1^{(k)}$, $c_2^{(k)}$) relevant to the 
MILC lattice data employed have been computed in 
3-loop lattice perturbation theory and, with the corresponding $d_k$, compiled
in Refs.~\cite{latticealphas05,hpqcd08}. $m_q$-dependent contributions were 
removed by extrapolation, using data, 
and non-perturbative (NP) $m_q$-independent 
higher $D$ contributions treated as being dominated by $D=4$ gluon 
condensate terms, which were fitted and removed
independently for each $O_k$.
Data with lattice spacings $a\sim 0.18,\, 0.12$, and $0.09\ fm$ were employed.
At these scales it was necessary to fit at least one additional coefficient
in Eq.~(\ref{3loopPT})~\cite{latticealphas05}. More recently, 
new MILC ensembles
with $a\sim 0.15$ and $0.06\ fm$ became available and 
were incorporated into the updated analyses of Refs.~\cite{hpqcd08,mlms08}.
One very new $a\sim 0.045\ fm$ ensemble was also employed in
\cite{hpqcd08}. The new analyses thus involve data whose 
range of scales is greater and whose highest scale is larger (and hence more
perturbatively-dominated). The two re-analyses, moreover, differ
somewhat in their strategies, allowing for useful cross-checks.
First, the two analyses
employ a different choice of coupling, that of Ref.~\cite{hpqcd08}
leaving residual perturbative uncertainties in the conversion
from the $V$ to ${\overline{MS}}$ scheme, that of Ref.~\cite{mlms08}
leaving them in the effects of the truncated $\beta$ function, which
can be suppressed by focussing on finer lattices~\cite{mlms08}. 
Second, Ref.~\cite{hpqcd08}
performs an improved treatment of $m_q$-independent NP contributions,
fitting a range of $D\geq 4$ forms to data,
while Ref.~\cite{mlms08} restricts its attention
to observables where the corresponding $D=4$ contributions,
estimated using charmonium sum-rule input for
$\langle \alpha_s G^2\rangle$~\cite{newgcond4}, can be shown to be
small. Even with finer lattice scales, at least one additional coefficient in 
Eq.~(\ref{3loopPT}) must be fit. The resulting fitted
$\alpha_s$ provide an excellent representation of the scale
dependence of the $O_k$. The results of the two
re-analyses are in good agreement, and differ by only $\sim 1\sigma$
from the results of \cite{latticealphas05}. The results, run
to the $n_f=5$ scale $M_Z$, are shown in Table~\ref{tablelattice} 
for the three most perturbative and four least perturbative of the
$O_k$ studied in \cite{hpqcd08}. $W_{kl}$
is the $k\times l$ Wilson loop and $u_0=W_{11}^{1/4}$.
Also shown is a measure, $\delta_{D=4}$, of the expected importance
of $m_q$-independent NP contributions to $O_k$, relative
to the $D=0$ contribution of interest in the determination of $\alpha_s$.
$\delta_{D=4}$ is the percent shift in the scale dependence between
$a\sim 0.12$ and $a\sim 0.06\ fm$ resulting from first computing 
$O_k$ using raw simulation values for the relevant $W_{kl}$,
and then re-computing it after subtracting the known leading order 
$m_q$-independent $D=4$ contributions, estimated using charmonium 
sum rule input for $\langle \alpha_s G^2\rangle$~\cite{newgcond4}.
Sizable NP effects are thus expected for the $O_k$ in the lower half 
of the table. The fact that, after such large contributions are 
approximately fitted and removed, the resulting $\alpha_s(M_Z)$ 
are in such good agreement with those obtained by analyzing 
more $D=0$-dominated $O_k$ argues strongly for the reliability of 
the approach and gives even higher confidence in results based 
on the most UV-sensitive of the $O_k$, $log\left( W_{11}\right)$, where
the estimated $D=4$ subtraction is very small, producing a
shift of only $0.0001$ in $\alpha_s(M_Z)$~\cite{mlms08}.

%\begin{figure}
%  \includegraphics[height=.3\textheight]{golfer}
%  \caption{Picture to fixed height}
%\end{figure}

%%%%%%%%%%%%%%%%%%%%%%%%%%%%%%%%%%%%%%%%%%%%
%% SAMPLE TABLE
%%
%% Shows the use of \tablehead and \tablenote
%% macros
%%%%%%%%%%%%%%%%%%%%%%%%%%%%%%%%%%%%%%%%%%%%

\begin{table}
\begin{tabular}{|lllc|}
\hline
$\qquad O_k$&\ \ $\alpha_s(M_Z)$&\ \ $\alpha_s(M_Z)$&$\delta_{D=4}$\\
&\ (HPQCD)&\ \, (CSSM)&\\
\hline
$\log\left( W_{11}\right)$&$0.1185(8)\ \quad$&$0.1190(11)$\quad&$0.7\%$\\
$\log\left( W_{12}\right)$&$0.1185(8)\ \quad$&$0.1191(11)$\quad&$2.0\%$\\
$\log\left( W_{12}/u_0^6\right)$&$0.1183(7)\ \quad$&$0.1191(11)$\quad
&$5.2\%$\\
\hline
$log\left(W_{11}W_{22}/W_{12}^2\right)$&$0.1185(9)\ \quad$&$\ \ \ N/A$\quad
&$32\%$\\
$log\left(W_{23}/u_0^{10}\right)$&$0.1176(9)\ \quad$&$\ \ \ N/A$\quad
&$53\%$\\
$log\left(W_{14}/W_{23}\right)$&$0.1171(11)\quad$&$\ \ \ N/A$\quad
&$79\%$\\
$log\left(W_{11}W_{23}/W_{12}W_{13}\right)$&$0.1174(9)\ \quad$&$\ \ \ N/A$
\quad&$92\%$\\
\hline
\end{tabular}
\caption{$\alpha_s(M_Z)$ and the shift $\delta_{D=4}$
induced by the $D=4$ $m_q$-independent NP correction with 
charmonium sum rule input for $\langle \alpha_s G^2\rangle$}
\label{tablelattice}
\end{table}

%\begin{table}
%\begin{tabular}{lrrrr}
%\hline
%  & \tablehead{1}{r}{b}{Single\\outlet}
%  & \tablehead{1}{r}{b}{Small\tablenote{2-9 retail outlets}\\multiple}
%  & \tablehead{1}{r}{b}{Large\\multiple}
%  & \tablehead{1}{r}{b}{Total}   \\
%\hline
%1982 & 98 & 129 & 620    & 847\\
%1987 & 138 & 176 & 1000  & 1314\\
%1991 & 173 & 248 & 1230  & 1651\\
%1998\tablenote{predicted} & 200 & 300 & 1500  & 2000\\
%\hline
%\end{tabular}
%\caption{Average turnover per shop: by type
%  of retail organisation}
%\label{tab:a}
%\end{table}

In the SM, with $\Gamma^{had}_{V/A;ud}$ the $\tau$ width to hadrons through 
the $I=1$ V or A current, $\Gamma_e$ the $\tau$ electronic width, 
$y_\tau =s/m_\tau^2$, and $S_{EW}$ a known short-distance EW correction, 
$R_{V/A;ud}=\Gamma^{had}_{V/A;ud}/\Gamma_e$ is
related to the spectral functions $\rho^{(J)}_{V/A;ud}(s)$ of the
spin $J$ scalar correlators, $\Pi_{V/A;ud}^{(J)}(s)$, 
of the V/A current-current two-point functions by~\cite{tsai}
\begin{equation}
dR_{V/A;ud}/dy_\tau = 12\pi^2S_{EW} \vert V_{ud}\vert^2
\left[ w_{00}(y_\tau )\rho^{(0+1)}_{V/A;ud}(s)-w_L(y_\tau )
\rho_{V/A;ud}^{(0)}(s)
\right] ,
\label{taubasic}\end{equation}
where $w_{00}(y)=(1-y)^2(1+2y)$, $w_L(y)=2y(1-y)^2$ and, 
up to $O\left[ (m_d\pm m_u)^2\right]$ corrections,
$\rho^{(0)}_{V;ud}(s)=0$ and $\rho_{A;ud}^{(0)}(s)=2f_\pi^2\delta (s-m_\pi^2)$.
$\rho^{(0+1)}_{V/A;ud}(s)$ is thus accessible from experimental results
for $dR_{V/A;ud}/dy_\tau$~\cite{alephud05,opalud99}. The corresponding 
correlator combination satisfies, for any $s_0$ and any analytic $w(s)$, 
the FESR relation
\begin{equation}
\int_0^{s_0}w(s)\, \rho^{(0+1)}_{V/A;ud}(s)\, ds\, =\, -{\frac{1}{2\pi i}}
\oint_{\vert s\vert =s_0}w(s)\, \Pi^{(0+1)}_{V/A;ud} (s)\, ds\ ,
\label{basicfesr}
\end{equation}
where the OPE can be employed on the RHS for large enough $s_0$.
For typical weights $w(s)$ and $s_0$ above $\sim 2\ {\rm GeV}^2$,
$\left[ \Pi^{(0+1)}_{V/A;ud} \right]_{OPE}$ is strongly $D=0$ dominated,
hence largely determined by $\alpha_s$. 
Use of polynomial weights, $w(y)$, with $y=s/s_0$, helps in quantifying
higher $D$ contributions, most of which must be fit to data,
since (with $N$ the degree of $w(y)$),
(i) up to corrections of $O\left(\alpha_s^2\right)$, the OPE series
terminates at $D=2N+2$, and (ii) integrated OPE
contributions with $D=2k+2$ scale as $1/s_0^k$, allowing contributions 
with different $D$ to be separated via their differing $s_0$-dependences.

Earlier $\tau$ decay determinations were based on 
combined analyses of the $s_0=m_\tau^2$, $km=00$, $10$, $11$, $12$, $13$, 
$w_{km}(y)=w_{00}(y)\, (1-y)^ky^m$ ``spectral weight FESRs''
(see, e.g., Refs.~\cite{davieretal08,alephud05}).
The most recent versions~\cite{davieretal08,bck08} employ the 5-loop $D=0$ 
V/A Adler function result~\cite{bck08}, as do subsequent 
studies~\cite{cgp08,jb08,kmty08,narison09,menke09}.
The $w_{km}$ analysis relies crucially on the additional
non-trivial assumption that $D=10,\cdots ,16$ contributions, 
each in principle present for one or more of the $w_{kl}$ employed, can, 
in all cases, be safely neglected, an assumption 
of potential relevance since a $\sim 1\%$
determination of $\alpha_s(M_Z)$ requires control of $D>4$
NP contributions to $\lesssim 0.5\%$ of the leading
$D=0$ term. Tests of this assumption were performed 
in Ref.~\cite{kmty08} by (i) studying 
the match between the OPE integrals, evaluated using fitted OPE parameters, 
and the corresponding experimental weighted spectral integrals as
a function of $s_0$, and (ii) using the same data and
fitted OPE parameters as input to FESRs for different
$w(y)$ involving the same set of OPE parameters. 
It was found that,
in the window $\sim 2\ {\rm GeV}^2<s_0\leq m_\tau^2$,
the match between the optimized OPE integrals and experimental
spectral integrals generated by the ALEPH data and fits
is typically poor, not just for the $w_{kl}$ employed in the ALEPH 
analysis, but also for other degree $\leq 3$ weights, which depend 
only on the $D=0,4,6,8$ OPE parameters included in the ALEPH fit. 
Similar problems, albeit
with somewhat reduced OPE-spectral integral discrepancies,
are also found for the OPAL data and fit parameter set.
Refs.~\cite{kmty08} also performed analyses based
on alternate weights, $w_N(y)=1-{\frac{N}{N-1}}y+{\frac{1}{N-1}}y^N$,
designed to suppress $D=2N+2$ contributions relative to the
leading $D=0$ terms, and hence optimize the determination of 
$\alpha_s$. It was found that
(i) the fits for $\alpha_s$ obtained using different $w_N(y)$, and also 
analyzing separately the V, A and V+A channels, are all in excellent agreement;
(ii) the impact of the $D>4$ OPE contributions is, as intended,
small; and
(iii) unlike the situation found using the ALEPH and OPAL fits,
the $w_N$ FESR fit parameter set produces OPE spectral integral 
results which match the corresponding spectral integrals within experimental 
errors for other degree $\leq 3$ weights (including the kinematic weight
$w_{00}$) over the whole of the $s_0$ window noted above.
One should bear in mind 
that, in terms of its size relative to the crucial $D=0$ term, 
it is a factor of between $7$ and 
$814$ times safer to neglect $D>8$ contributions in the 
$w_N$ analyses than it was in the higher $w_{kl}$ FESRs
of the ALEPH and OPAL analyses~\cite{kmty08}. In view of the fact that 
(i) the older analyses, which should produce results in agreement
with those of the corresponding $w_N$ analyses when using the
same data, instead produce significantly larger $\alpha_s$, and 
(ii) the results of the old analyses, considered at lower $s_0$, 
produce optimized OPE integrals not in agreement within errors 
with the corresponding experimental spectral integrals, 
and, moreover, significantly inferior to the
matches obtained using the $\{ w_N\}$ analysis fit parameters
(see, e.g., the Figures in Refs.~\cite{kmty08}), it seems
clear that the results of the $\{ w_N\}$ analysis should
be taken to supercede those of the earlier combined $w_{kl}$
analyses. The favored $\tau$ decay result for $\alpha_s$ is thus
\begin{equation}
\alpha_s(M_Z)=0.1187(16)\ ,
\end{equation}
in excellent agreement with the lattice determination.

We conclude with a few comments on other recent results for
$\alpha_s$ from hadronic $\tau$ decays. First, note that
Refs.~\cite{bck08,davieretal08,menke09} employ as input for their 
$D=6,8$ OPE parameter values, the results obtained in either the
2005 or 2008 ALEPH combined $s_0=m_\tau^2$ spectral weight FESR 
analysis. They thus lead to $s_0$-dependent OPE integrals
which do not match the corresponding spectral integrals within
experimental errors, and whose matches are inferior to those produced by
the OPE parameters obtained from the $w_N$ FESR analyses.
Ref.~\cite{jb08} (whose results also lead to an OPE-spectral
integral mismatch~\cite{kmty08}, this time resulting
from the use of a different set of assumed values for
the required $D=6,8$ input~\cite{kmty08}) however, raises an interesting
question about the relative reliability of the FOPT and
CIPT prescriptions for evaluating the truncated $D=0$ series,
one in need of, and undergoing, further investigation. Also relevant
in this regard is the observation of Ref.~\cite{menke09},
which shows a larger-than-previously-anticipated FOPT
uncertainty associated with the dependence of the truncated
FOPT result on the point on the OPE contour chosen as the 
fixed scale.

%\begin{figure}
%\unitlength1cm
%\rotatebox{270}{\mbox{
%  \includegraphics[height=10.0cm]{aleph05_veccomparisons_tau08.ps}
%}}
%  \caption{Comparison of fit qualities for degree $\leq 3$
%weight FESRs for the fits of Refs.~\cite{alephud05} and \cite{kmty08}}
%\end{figure}

%%%%%%%%%%%%%%%%%%%%%%%%%%%%%%%%%%%%%%%%%%%%%%%%
%% BACKMATTER
%%%%%%%%%%%%%%%%%%%%%%%%%%%%%%%%%%%%%%%%%%%%%%%%

\begin{theacknowledgments}
The support of the CSSM, University of Adelaide, and 
the Natural Sciences and
Engineering Research Council of Canada are gratefully acknowledged.
\end{theacknowledgments}

\end{document}